\documentclass[numbers,copyright,creativecommons]{eptcs}
\providecommand{\event}{2007} 
\providecommand{\volume}{??}
\providecommand{\anno}{20??}
\providecommand{\firstpage}{1}
\providecommand{\eid}{??}
\usepackage{breakurl}        
\usepackage{url}
\usepackage{amssymb}
\usepackage{amsthm}
\usepackage{multirow}
\usepackage{graphicx}
\usepackage{amssymb}
\usepackage{enumerate}
\usepackage{delarray} 
\usepackage{color}
\usepackage{amsfonts,amssymb,amsmath}
\usepackage{latexsym,amsfonts,amssymb}
\newcommand{\LemName}[1]{}

\newcommand{\Eqn}[1]{(\ref{#1})}
\newcommand{\figg}[1]{Fig.~\ref{#1}}

\newenvironment{Fig}{\begin{figure}[h]\small\rule{\textwidth}{.02cm}}{\rule{\textwidth}{.02cm}\end{figure}}
\newcommand{\st}{~{\scriptscriptstyle ^\bullet}~}

\newcommand{\Lift}[1]{[{#1}]}

\newcommand{\define}{~ \hat{=} ~}

\newcommand{\Pred}{\textit{Pred}}

\newcommand{\EE}[1]{ {\cal E}{#1}}

\newcommand{\cubes}{\mbox{\sf cubes}}
\newcommand{\cubed}{\mbox{\sf cubed}}

\newcommand{\Gets}{:=\,}
\newcommand{\PC}[1]{\mathbin{\makebox[0em]{~}_{#1}\oplus}}
\newcommand{\Ch}{\ [\!] \ }

\newcommand{\Abort}{\mbox{\sf abort}}
\newcommand{\Skip}{\mbox{\sf skip}}
\newcommand{\If}{\mbox{\sf if\ }}

\newcommand{\Thn}{\mbox{\sf \ then\ }}
\newcommand{\Els}{\mbox{\sf \ else\ }}

\newcommand{\Do}{\mbox{\sf do\ }}
\newcommand{\Od}{\mbox{\sf \ od}}

\newcommand{\Wp}{\mbox{\sf wp}}

\newcommand{\DWp}{\mbox{\sf wp}}

\newcommand{\prog}{\textit{Prog}}

\newcommand{\Min}{~\sqcap~}

\newcommand{\Max}{~\sqcup~}

\newcommand{\pGCL}{\mbox{pGCL}}
\newcommand{\GCL}{\mbox{GCL}}

\newcommand{\Prog}{\textit{Prog}}

\newcommand{\IE}{\textrm{i.e.}}

\newcommand{\CompNY}[3]{(#1#2 \Spot #3)}
\newcommand{\Spot}{\mathrel{\mbox{\boldmath$\cdot$}}}

\newcommand{\Ref}{\sqsubseteq}

\newcommand{\Defs}{\mathrel{\hat{=}}}

\newcommand{\Wide}[1]{~~~#1~~~}
\newcommand{\Sec}[1]{Sec.~\ref{#1}}
\newtheorem{Lems}{Lemma} 
\newcommand{\Lem}[1]{Lem.~\ref{#1}}
\newcommand{\Cor}[1]{Cor.~\ref{#1}}

\newtheorem{Defns}{Definition}
\newtheorem{Corl}{Corollary}
\newcommand{\Def}[1]{Def.~\ref{#1}}
\newcommand{\Proof}{\noindent{\bf Proof}: \quad}
\newenvironment{Reason}{\begin{tabbing}\hspace{4em}\=\kill}{\end{tabbing}\vspace{-2.5ex}}

\newcommand{\Step}[2]{#1 \> $\begin{array}[t]{ll}#2\end{array}$ \\}
\newcommand{\StepR}[3]{#1 \> $\begin{array}[t]{ll}#3\end{array}$ \` {\RF #2} \\}

\newcommand{\RF}{\small}

 \DeclareMathSymbol\RRightarrow{\mathrel}{AMSa}{"56}
 \DeclareMathSymbol\LLeftarrow{\mathrel}{AMSa}{"57}

 \newcommand{\bag}[2]{{[[\ #1\ ]]}^{#2}}

 \newcommand{\comp}{\mathbin{\raise 0.6ex\hbox{\oalign{\hfil$\scriptscriptstyle
     \mathrm{o}$\hfil\cr\hfil$\scriptscriptstyle\mathrm{9}$\hfil}}}}
\newcommand{\pp}[1]{\mathbin{\makebox[0em]{~}_{#1}\oplus}}

\title{An expectation transformer approach to predicate abstraction and data independence\\ for probabilistic programs\thanks{The authors acknowledge support from (I) The Australian Commonwealth Endeavour International
Postgraduate Research Scholarship (E-IPRS) Fund, and (II) The  Australian Research Council (ARC) Grant Number DP0879529.}}
\author{Ukachukwu Ndukwu
and AK McIver
\institute{Deptartment of Computing, Macquarie University, NSW 2109 Australia.}
\email{\{ukndukwu,anabel\}@science.mq.edu.au}
}
\def\titlerunning{An expectation transformer approach to predicate abstraction}
\def\authorrunning{Uk Ndukwu \ and AK McIver}

\begin{document}
\maketitle

\begin{abstract}
In this paper we revisit the well-known technique of predicate abstraction to characterise performance attributes of system models incorporating
probability.\ We recast the theory using expectation transformers \cite{ARP}, and identify transformer properties which correspond to abstractions that yield nevertheless exact bound on the performance of infinite state probabilistic systems.\ In addition, we extend the developed technique to the special case of  ``data independent'' programs \cite{Wolper86} incorporating probability.\ Finally, we demonstrate the subtleness of the extended technique by using the PRISM model checking tool \cite{PRISM} to analyse an infinite state protocol, obtaining exact bounds on its performance.


\end{abstract}

\setcounter{section}{0}

\section{Introduction}\label{s0154}
Automated analysis of infinite (very large) state systems often relies on abstractions which summarise the essential behaviour as a finite state ``anti-refinement" in such a way as to guarantee the desired properties (if indeed they hold). Typically, however, abstractions introduce nondeterminism, and in a probabilistic system this can lead to a high degree of imprecision in the estimated probabilistic properties. The choice of abstraction therefore is critical; some approaches to finding the right one use ``abstraction refinement", sometimes relying on counterexamples of failed attempts to obtain incremental improvements \cite{CEGAR}.

In this paper we revisit the technique of ``predicate abstraction" from the perspective of ``expectation
transformers".\ \emph{Predicate abstraction} refers to the notion of approximating a system using a given set of predicates: states are grouped together according to the predicates they satisfy (in the given set), and the system is abstracted by tracking only the transformations expressible in the induced equivalence classes.\ \emph{Expectation transformers} \cite{ARP} is a generalisation to probabilistic systems of Hoare/Dijkstra-style semantic reasoning \cite{Dijkstra76} --- predicates are replaced by real-valued functions of the state.\ The approach is equivalent to operational models of programming based on Markov-Decision Processes, but results in a convenient proof system for verifying general properties of probabilistic programs.

In particular we are able to characterise, using expectation transformers, a simple criterion for when an abstraction gives \emph{exact} quantitative analysis for probabilistic properties.\ The criterion is sufficient to identify when predicate abstraction introduces no additional nondeterminism.\ A typical class of programs where this is effective is the so-called class of ``data independent" programs \cite{Wolper86}.\ A program is \emph{data independent} whenever its control structure does not depend on the precise values of the data.\ Wolper \cite{Wolper86} first identified this as a class of interesting programs amenable to verification via model checking \cite{CGP99}.\ In addition to Wolper's idea, we consider the notion of probabilistic data independence where the probabilistic choice cannot be dependent on the data.\ In general, the idea we propose here is aimed at constructing abstractions which result in no loss of information especially when probability plays a crucial role in the performance analysis of infinite state system models.\ Such abstractions are said to be ``information-preserving'' since they suffice as exact representations of their original systems.

Using the expectation transformer approach we prove the ``folk theorem'' (see \cite{FM86}) for probabilistic systems: that data independent programs can be treated with predicate abstraction yielding exact results on threshold properties such as ``the probability that a set of states has been reached in at most $k$ steps".

In particular our contributions in this paper are:
\begin{enumerate}
\item [(i)] The development of a technique which permits the application of predicate abstraction to probabilistic programs using expectation transformers;

\item [(ii)] An establishment of a criterion for identifying abstractions which do not lose information;

\item [(iii)] We show how the developed technique and criterion can be applied to data independent programs
    especially when probability plays a crucial role;

\item [(iv)] And finally, a demonstration of the technique on a case study of a system with potential infinite state behaviour.
\end{enumerate}

 This paper is structured as follows: In \Sec{ppt} we summarise the expectation transformer semantics for probabilistic programs, \Sec{AET} is the development of the technique for predicate abstraction using expectation transformers.\ In \Sec{PDI} we show how the technique can be applied to identify when  predicate abstraction yields exact thresholds for infinite state systems; we then explore the special treatment of data independent programs.\ In \Sec{DC} we illustrate the technique by model checking the Rabin's choice coordination problem (also known as the distributed consensus) \cite{Rabin82}; this is a protocol which  has the potential to require unbounded resources on its performance and therefore cannot be verified directly with a model-checking approach. However the theory of \Sec{PDI} shows that the results we obtain using its information-preserving abstraction are nevertheless exact interpretations of its performance.

\subsection{Summary of notation}
Function application
is represented by a dot, as in $f\cdot x$ (rather than $f(x)$).
We use an abstract state space $S$.
 Given predicate $\Pred$ we write
$\Lift{\Pred}$ for the \emph{characteristic} function mapping states satisfying $\Pred$ to
$1$ and to $0$ otherwise, punning $1$  and $0$ with ``True" and ``False" respectively.\ Whenever $e, e'$ are real-valued functions  over $S$ we write $e+e'$, $e\Max e'$, $e \Min e'$ for the pointwise addition, maximum and minimum. Moreover $\alpha {\times} e$ is $e$ scaled by the real $\alpha$.

For commutative operator $\odot$, we use  $\CompNY{\odot}{x \in X}{f\cdot x}$ for the comprehension which applies $\odot$ between all instances $f\cdot x$ as $x$ ranges over $X$.\ For example, $(\Max x \in [0, 1] \cdot x^{2})$ gives the maximum value of $x^{2}$ as $x$ ranges over the closed interval $[0, 1]$.

\section{Probabilistic program semantics and expectation transformers}\label{ppt}

When programs incorporate probability,  their properties can no longer be guaranteed
``with certainty", but only
 ``up to some probability".
For example the program
\begin{equation}\label{e1524}
\textit{inc} \Wide{\Defs} x \Gets x/2 ~\PC{1/2}~ x{+}1~,
\end{equation}

sets the integer-valued variable $x$ to $x/2$ (the result of the integer division) only with probability $1/2$ --- in practice
this means that
if the statement \Eqn{e1524} were executed a large number of times, and the number of times that $x$ was halved or increased  tabulated,
roughly $1/2$ of them would record $x$ as having been halved
 (up to well-known statistical confidence \cite{GW86}).

The probabilistic guarded command language $\pGCL$ \cite{ARP}  and its associated \emph{quantitative logic}
were developed to express such programs and to derive their probabilistic
properties by extending the classical assertional style of programming.\ Programs in the $\pGCL$
are modeled (operationally) as functions (or transitions) which
map \emph{initial states} in $S$
 to (sets of) probability distributions over \emph{final states} ---
 the program at \Eqn{e1524}
for instance has a single transition
 which
 maps any initial state $x=k_0$ to a (single) final distribution; we represent that distribution as a
 function $d$, evaluating to $1/2$ when $x=k_0/2$ or $x=k_0{+}1$.

Since properties are now quantitative we express them via  a logic of
\emph{real-valued functions}, or \emph{expectations}.
   For example,
 the property ``if the initial state satisfies $x=0\lor x=2$, then the final value of $x$ is $1$ with probability $1/2$"
can be expressed as the \emph{expected value} of
the function $[x=1]$
with respect to $d$, which evaluates to
$1/2 \times 1 + 1/2 \times 0 = 1/2$, when $x$ is initially $2$ for example.

Direct appeal to the operational semantics quickly becomes
impractical for all but the simplest programs --- better is the equivalent transformer-style semantics which is
obtained by
rationalising the above calculation in terms of expected values rather than transitions, and the explanation
runs as follows.  Writing ${\cal E} S$ for the set of all (non-normalised) functions from $S$ to the interval $[0,1]$,
which we call the set of \emph{expectations}, we say that
the expectation $[x=1]$ has been transformed to the expectation $[x=0 \lor x=2]/2$
 by the program \textit{inc} (\ref{e1524}) above so that they are in the relation
 ``$1/2$ is the
 expected value of $[x=1]$ with respect to \textit{inc}'s result distribution whenever $x$ is initially either $0$ or $2$".
More generally given a program $\prog$, an expectation $e$ in ${\cal E} S$ and a state $s \in S$,
we define ${\sf wp}.\prog.e.s$ to be the expected value of $e$ with respect to the
result distribution of program $\prog$ if executed initially from state $s$.\ We say that ${\sf wp}.\prog$ is the \emph{expectation transformer} relative to $\prog$.\ In our example that allows us to write
 \[
 [x=0 \lor x=2]/2 \Wide{=} {\sf wp}.(x \Gets x/2 \PC{1/2} x \Gets x{+}1).[x=1]~.
 \]
    In the case that
\emph{nondeterminism} is present,  execution of $\prog$ results in a \emph{set}
of possible distributions and we modify the definition
of ${\sf wp}$ to take account of this --- indeed we define ${\sf wp}.\prog.E.s$
so that it  delivers the
\emph{least}-expected value with respect to all distributions in the result set.
The transformers \cite{ARP} give rise to a complete characterisation
of probabilistic programs with nondeterminism, and they are sufficient to
express many performance-style properties, including the probability that an event
occurs, the expected time that it occurs, and long-run
average of the number of times it occurs over many repeated executions of the system.


In \figg{f0604} we set out the semantics for the $\pGCL$, a variation of Dijkstra's
$\GCL$ with the addition of  probabilistic choice.  All the programming
 features have been defined previously
elsewhere, and (apart from probabilistic choice) have interpretations which are merely
adapted to the real-valued context. For example,
nondeterminism, as explained above, is interpreted \emph{demonically}
and  can be thought of as being resolved by a ``minimal-seeking demon",
providing
guarantees on all program behaviour, such as is expected for total correctness.
 \emph{Probabilistic choice}, on the other hand, selects the operands
at random with weightings determined by the probability parameter $p$.\ Iteration is defined by a least fixed point of a monotone expectation-to-expectation function.\footnote{Well-definedness is guaranteed by, for example, restricting the expectations to lie in the real interval $[0,1]$ or to complete the reals with $\infty$. These issues have been discussed elsewhere \cite{ARP}.}

\begin{Fig}{
{\small
\[
\begin{array}{ll}
\textit{skip} &\DWp.\Skip.E \define E~,\\
\textit{abort} & \DWp.\Abort.E \define 0~,\\
\textit{assignment} &\DWp.(x \Gets f).E \define E[x \Gets f]~,\\
\textit{sequential composition} &\DWp.(r\comp r').E \define \DWp.r.(\DWp.r'.E)~,\\
\textit{probabilistic choice} &\DWp.(r \PC p r').E \define p \times \Wp.r.E + ({1} \mathord- p)\times \Wp.r'.E~,\\
\textit{nondeterministic choice} &\DWp.(r \Ch r').E \define  \DWp.r.E \Min \DWp.r'.E~,\\
\textit{Boolean choice} &\DWp.(\If G \Thn r \Els r').E \define \Lift{G} \times \DWp.r.E + \Lift{\neg G} \times \DWp.r'.E~,\\
\textit{Iteration} &\DWp.(\Do ~ G \rightarrow r ~ \Od).E \define (\mu X \st  [\neg G]{\times} E +[G]{\times} \DWp.r.X)~.
\end{array}
\]
$E$ is an expectation in $\EE{S}$, and  $f$ is a function of the state,
and $\Min$ is pointwise minimum.\ The real $p$ is restricted to lie between $0$ and $1$, and the term ($\mu X \dots $) refers to the least fixed point with respect to $\leq$, which we lift to real-valued functions.\ Commands are ordered using \emph{refinement}, so that more refined programs
improve probabilistic results, thus $P \Ref Q \Wide{\textit{iff}} \CompNY{\forall}{E \in \EE{S}}{\Wp.P.E \leq \Wp.Q.E};$
note also that the {\it monotone} property of wp is such that if $E \le E^{'}  \mathrm{then} \ {\Wp.P.E \leq \Wp.P.E^{'}}$,
where $P,Q$ are program commands and $E, E^{'}$ are expectations.
}}

\caption{Structural definitions of $\DWp$ for  the $\pGCL$.}\label{f0604}
\end{Fig}


We end this section with a discussion of  a simple performance property: a probabilistic analysis of the number of iterations until termination.  Given a loop $\Do ~ G \rightarrow \Prog \Od$ which executes the program $\Prog$ until $G$ becomes false, we can compute the probability that the loop has executed no more than $k$ times on termination as:

\begin{equation}\label{e0933}
\Wp.\Do ~ G \rightarrow \Prog; n\Gets n{+}1 \Od.[n\leq k]~,
\end{equation}
where $n$ is a fresh variable, not occurring in $\Prog$. Informally, if $n$ is initialised to $0$ before the execution of the loop and is incremented after each execution of $\Prog$, this expresses the (minimum) probability that its value on exiting the loop does not exceed $k$.\ When no nondeterminism is present the expression in \Eqn{e0933} computes an exact bound for expected performance; when it is present it computes the greatest lower bound.\ However upper bounds can be calculated using a maximum interpretation of nondeterminism but we do not discuss that interpretation here.

In this section we have summarised an expectation transformer approach to probabilistic semantics. In many cases, especially for performance, the exact analysis of the system in this style is impractical; an alternative to calculation is model checking, however this is not viable for very large or infinite systems. Predicate abstraction is a popular approach to approximating such programs, and in the next section we develop the expectation transformer approach to predicate abstraction for probabilistic programs.

\section{Abstract expectation transformers}\label{AET}

Predicate abstraction is a standard technique for defining abstractions of transition systems.\ In this section we will show how to define it for probabilistic programs using expectation transformers.\ The approach is inspired by Ball's formalisation of predicate abstraction for standard sequential programs  using weakest precondition semantics \cite{Ball}.

Let $\Phi$ be a (finite) set of predicates over the state space $S$. The standard predicate abstraction over $\Phi$ is induced by the equivalence class:
\[
s \sim_\Phi s' ~\Wide{\textit{iff}} ~ (\forall \phi\in \Phi \cdot \phi.s = \phi.s')~.
\]
Given a transition system $T$ over $S$, the abstract transition system $T/\!\!\sim_\Phi$ takes the equivalence classes given by $S/\!\sim_\Phi$ as the state space, and their transitions $\hat{s} \rightarrow \hat{t}$ in $T/\!\sim_\Phi$ provided that there exists a transition $s \rightarrow t$ in $T$.\ The probabilistic generalisation is somewhat more complicated to define.\ On the other hand the expectation transformer semantics characterises operational behaviour, and the approach we take here is to define the abstract transition system over $S/\Phi$ using a generalisation of Ball's idea.

Let $\cubes.\Phi$ be the (finite) set of (non trivial) minimal predicates formed by taking negations and conjunctions of predicates in $\Phi$. The set $\cubes.\Phi$ corresponds to the (set of) equivalence classes $S/\!\!\sim_\Phi$, and represents the abstract state space of the abstraction induced by $\Phi$.
Let $\cubed_\Phi : \EE S \rightarrow \EE S$ be defined

\begin{equation}\label{e0722}
\cubed_\Phi.e \Wide{\Defs}  \CompNY{\Max}{c\in \cubes.\Phi} {\CompNY{\Max}{\lambda [c] \leq e}  {\lambda [c]}}~.
\end{equation}
We note that $\cubed_\Phi.e$ is unique and would usually be a linear combination of the elements of $\cubes.\Phi$, hence making it the sum of scaled cubes over the latter.\ Consequently, $\cubed.e$ is the weakest approximation of $e$ with respect to the granularity expressible by conjunctions, negations and disjunctions in $\Phi$.
We say that $e$ is \emph{cubed} relative to $\Phi$ exactly when $\mathord{e = \cubed_\Phi.e}$. Note that $\mathord{\cubed.\cubed.e = \cubed.e}$ and
that  sums, maxima and minima of cubed expressions are still cubes, \IE\
\[
\begin{array}{ll}
3(a)~~~ & \cubed.(\cubed.e + \cubed.e') \  =  \ (\cubed.e + \cubed.e')~, \\

3(b)~~~ &  \cubed.(\cubed.e \Max \cubed.e') = (\cubed.e \Max \cubed.e')~, \\

3(c)~~~ &  \cubed.(\cubed.e \Min \cubed.e') = (\cubed.e \Min \cubed.e') ~. \\

\end{array}
\]


\begin{Defns}\label{d1822}
Given a $\pGCL$ program $\Prog$, and a set of predicates $\Phi$, and expectation $e$ over $S$ we define the abstract weakest expectation relative to $\Phi$ as:
\[
\Wp_\Phi.\Prog.e \Wide{\Defs} \cubed_\Phi. \Wp.\Prog.e~.
\]
We write $\Prog_\Phi$ for the corresponding abstract program operating over the abstract system $S/\Phi$.\ This implies that
$Prog_{\Phi}$ is determined by $\Wp_\Phi.\Prog$.
\end{Defns}

\begin{figure}
\begin{center}
\includegraphics[scale=0.5]{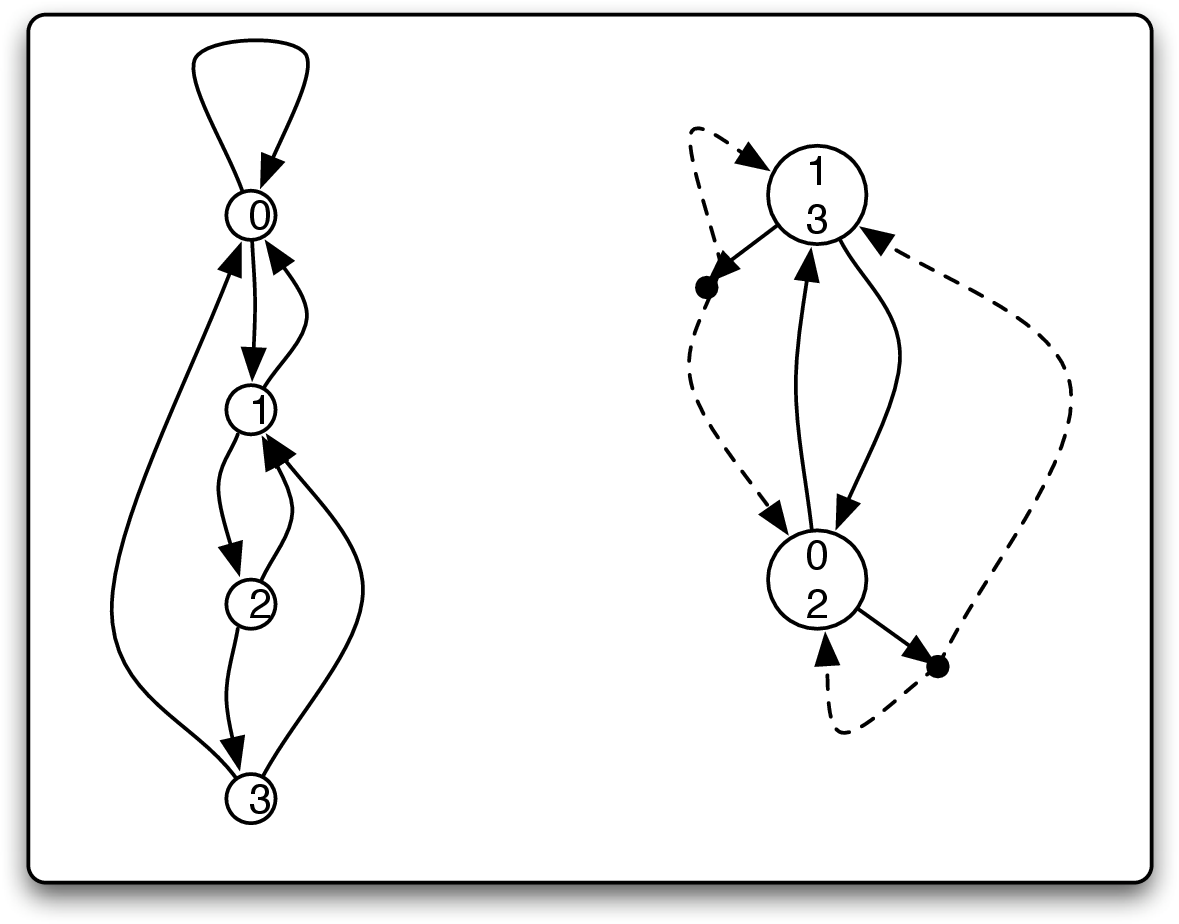}
\end{center}

\small{The transition system on the left represents the program \textit{inc} over the state space defined by $0 \leq x < 4$, using arithmetic modulo 4. Each solid black arrow occurs with probability $1/2$. The transition system on the right is the abstraction based on $\Phi = \{x=0 \lor x=2\}$. Here we can see non determinism (indicated by dotted lines) is introduced after any transition which divides the value of $x$ by $2$.}
\caption{The transition system for $\textit{inc}$ and an abstraction.}\label{f1414}
\end{figure}

As an example, consider the program  \textit{inc} at \Eqn{e1524} operating under arithmetic modulo $4$. The underlying state space is defined by $0 \leq x < 4$; consider now the set  $\Phi$ consisting of the single predicate $x=0 \lor x=2$; the set of cubes $\cubes.\Phi \Defs \{(x=0 \lor x=2), (x=1 \lor x=3)\}$, implying that the induced predicate abstraction has two states. We can see now that
\[
\begin{array}{ll}
&\Wp_\Phi.\textit{inc}.[x=1 \lor x=3] = [x=0\lor x=2]/2~, \\
 \textit{and}~~~ &  \Wp_\Phi.\textit{inc}.[x=0 \lor x=2] = [x=1\lor x=3]/2~,
 \end{array}
\]
which is consistent with the abstraction in \figg{f1414}, where each abstract state has a probability of at least $1/2$ of
being transformed to the other state, with the remaining probability being assigned to a nondeterministic update.

The next lemma sets out some properties of the abstract expectation transformer.

\begin{Lems}\label{l1541}
Let $\Prog, \Prog'$ be  programs, $\Phi, \Phi'$  sets of predicates, and $e, e'$ expectations and $\alpha$ a real. The following inequalities apply:
\[
\begin{array}{llll}
(1)~~~ & \Wp_\Phi.\Prog.e & \Wide{\leq}& \Wp.\Prog.e\\

(2)~~~ &  \Wp_{\Phi} . \Prog.e &\Wide{\leq}& \Wp_{\Phi \cup \Phi'}.\Prog.e\\

(3)~~~ & \Wp_\Phi.\Prog.e + \Wp_\Phi.\Prog.e' &\Wide{\leq}& \Wp_\Phi.\Prog.(e + e')  \\
(4)~~~ & \alpha{\times}\Wp_\Phi.\Prog.e  &\Wide{=}& \Wp_\Phi.\Prog.(\alpha{\times}e)  \\

(5) ~~~ &(\Wp_\Phi.\Prog.e{-}1)\Max 0)  &\Wide{\leq}& \Wp_\Phi.\Prog.((e{-}1)\Max 0)  \\

\end{array}
\]

\Proof
The inequalities and equalities all follow from arithmetic and \Def{d1822}.
\end{Lems}

\Lem{l1541} confirms our intuition that (1) the properties measured with respect to the abstraction are no more than with respect to the original program; (2) finer-grained abstractions give more accurate results and (3,4,5) $\Wp_\Phi.\Prog$ corresponds to a well-defined probabilistic transition system \cite{ARP}.

For standard transitions systems (with no probability) an abstract system $\Prog_\Phi$ is determined directly from the control structure and assignment statements. This corresponds to $\Wp_\Phi$ distributing through the program operators.  For probabilistic systems this is not the case. For example, the program $\textit{inc};\textit{inc}$ (with addition modulo $4$) we may compute $3[x=0 \lor x=2]/4 \leq \Wp_\Phi.(\textit{inc};\textit{inc}).[x=0 \lor x=2]$, whereas $[x=0 \lor x=2]/4 =\Wp_\Phi.\textit{inc}. (\Wp_\Phi.\textit{inc}).[x=0 \lor x=2]$, implying that the nondeterminism introduced at each abstract transition will increase the inaccuracy. Comparing with \figg{f1414} we see that nondeterminism is introduced at each abstract transition, and this could be resolved in the abstract system in such a way that there is only $1/4$ chance of returning to the initial abstract state.

 The following lemma shows that $\Wp_\Phi$ only distributes through nondeterminism, and only subdistributes through sequential composition and probabilistic choice.

\begin{Lems}\label{l1508}
Let $\Prog, \Prog'$ be  programs, $\Phi, \Phi'$  sets of predicates, and $e, e'$ expectations. The following inequalities apply:

\[
\begin{array}{llll}
(4)~~~ & \Wp_\Phi.(\Prog \Ch \Prog') .e  &\Wide{=}& \Wp_\Phi.\Prog.e  \Min \Wp_\Phi.\Prog'.e \\

(5)~~~ & \Wp_\Phi.\Prog . (\Wp_\Phi. \Prog') .e  &\Wide{\leq}& \Wp_\Phi.(\Prog ; \Prog') .e \\

(6)~~~ &  \Wp_\Phi.(\Prog) \PC{p} \Wp_\Phi. \Prog' .e  &\Wide{\leq}& \Wp_\Phi.(\Prog \PC{p} \Prog') .e \\
\end{array}
\]

\Proof
The inequalities and equalities all follow from arithmetic and \Def{d1822}.
\end{Lems}

\begin{figure}
\begin{center}
\includegraphics[scale=0.5]{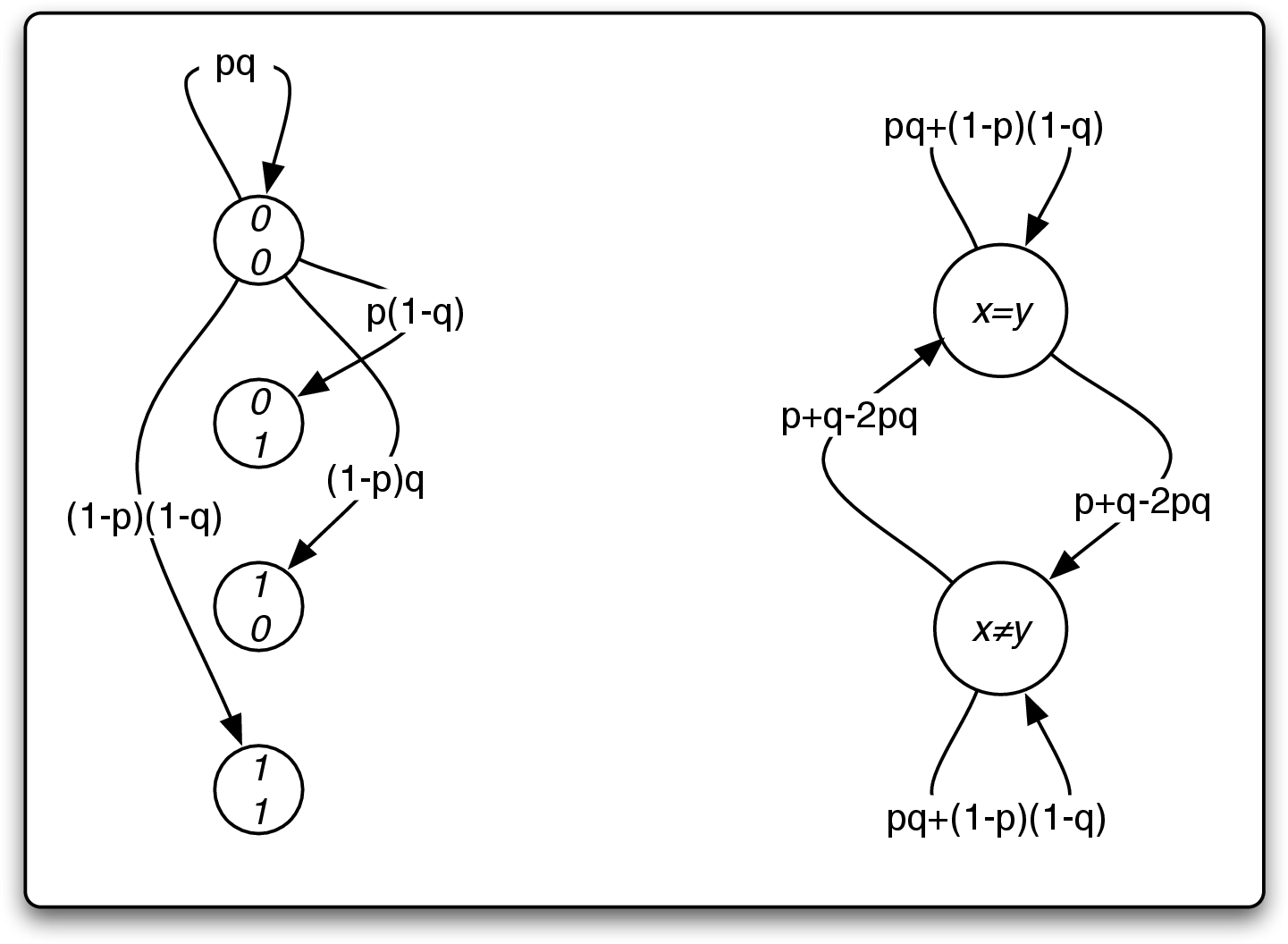}
\end{center}

\small{The transition system (labelled with probabilities) on the left represents the program \textit{twoFlip} over the state space defined by variables $x$ and $y$, each of which can take 0 or 1 value in the states ($x$,$y$). Each branch is executed with the probability that it occurs; only the transitions from $x=y=0$ are illustrated, with transitions from the remaining states similarly calculated.\ The transition system on the right represents the abstraction which only keeps track of whether $x$ and $y$ are equal or not. Since no nondeterminism is introduced, properties at that level of granularity can be accurately calculated using this abstraction.}
\caption{The transition system for $\textit{twoFlip}$ and an abstraction.}\label{f1716}
\end{figure}

\Lem{l1508} implies that whenever nondeterminism is introduced, the analysis of a program abstracted at each program statement could be too coarse to verify a desired quantitative threshold.\ This is not a problem when the abstraction
does not introduce nondeterminism. Consider the program
\begin{equation}\label{e1720}
\textit{twoFlip} \Wide{\Defs} x \Gets 0 \PC{p} x \Gets 1 ;  y \Gets 0 \PC{q} y \Gets 1~,
\end{equation}
and the set of predicates $\Phi \Defs \{x=y, x \neq y\}$.\ The resulting transition system over the state space defined by $x$ and $y$ is set out in \figg{f1716} together with the abstraction induced by this $\Phi$.

Observe how no nondeterminism has been introduced in this abstraction --- since indeed \\$\mathord{\Wp_\Phi.(twoFlip;twoFlip) = \Wp_\Phi.(twoFlip);\Wp_\Phi.(twoFlip)}$.\ Intuitively this tells us that properties which can be stated at the granularity of $\Phi$ can be computed accurately from its corresponding abstraction. In the next section we formalise our intuition using  expectation transformers.

\section{Information-preserving abstractions and expected time to terminate}\label{PDI}

In this section we introduce ``information-preserving" abstractions and study how they apply to the computation of exact bounds on performance-style properties of probabilistic programs.

As we saw above, an abstraction which does not introduce nondeterminism preserves the exact behaviour of the program at the granularity of the chosen set of predicates.\ Programs which do not exhibit nondeterminism or aborting behaviours satisfy the special properties that:
\[
\begin{array}{lll}
~~~  \Wp_\Phi.\Prog .(e + e^{'}) &\Wide{=}& \Wp_\Phi.\Prog.e  + \Wp_\Phi.\Prog.e^{'} \\

~~~  \Wp_\Phi.\Prog .(1 - e)  &\Wide{=}& 1 - \Wp_\Phi.\Prog .e \\

\end{array}
\]
for any {\it deterministic} pGCL program command $\Prog$, set of predicate $\Phi$, and  expectations $e, e'$.\ The next definition formalises that idea in terms of expectation transformers.

\begin{Defns}\label{d0636}
Given a deterministic program $\Prog$ and a set of predicates $\Phi$, we say that the abstraction induced by $\Phi$ is
\emph{information-preserving} if:
\[
\Wp_\Phi.\Prog.[c] \Wide{=} \Wp.\Prog.[c]~,
\]
for all $c\in \cubes.\Phi$.
\end{Defns}

To see \Def{d0636} in action, observe that

\begin{Reason}
\Step{}
{\Wp.\textit{inc}.[x=0 \lor x=2]}
\Step{$ =$}
{ [x=0 \lor x=3]/2 {+} [x=1]}
\Step{$ {\neq}$}
{[x=1 \lor x=3]/2}
\Step{$=$}
{ \Wp_\Phi.\textit{inc}.[x=0 \lor x=2] ~, }
\end{Reason}
implying the introduction of a nondeterministic branch at the abstract state corresponding to $x=1 \lor x=3$.

A more efficient way to check for information-preservation is simply to check that $\Wp.\Prog.[\phi]$ is cubed for all $\phi \in \Phi$; the next lemma shows that this is sound.

\begin{Lems}\label{l1307}
Let $\Prog$ be a deterministic (probabilistic) program, and let $\Phi$ be a set of predicates.
If $\Wp.\Prog.[\phi]$ is cubed for all $\phi \in \Phi$ then the abstraction induced by $\Phi$ is information-preserving.

\Proof
We need to show that $\Wp.\Prog.[c] = \Wp_\Phi.\Prog.[c]$ for all $c \in \cubed.\Phi$. Note that each such $c$ is generated from negations and conjunctions, so all we need show is that for predicates $\psi, \psi'$ such that $\Wp.\Prog.[\psi]$ and $\Wp.\Prog.[\psi']$ are cubed, then so too are $\Wp.\Prog.(1-[\psi])$ and $\Wp.\Prog.[\psi \land \psi']$.

The result follows since $\Wp.\Prog.[\psi \land \psi'] = (\Wp.\Prog.[\psi] + \Wp.\Prog.[\psi]' - 1)\Max 0$, and
$\Wp.\Prog.(1-[\psi]) = 1 - \Wp.\Prog.[\psi]$, and the fact that sums and inverses of cubed expressions are still cubed.
\end{Lems}

As mentioned above, a key characterising property of information-preserving abstractions is that they generate no new nondeterminism.
A probabilistic program exhibits no (demonic) nondeterminism if its expectation transformer semantics distributes addition.\ The next lemma shows this for information-preserving abstractions.

\begin{Lems}\label{l0723}
Let $\Prog$ be a deterministic (probabilistic) program, and let $\Phi$ be a set of predicates inducing an information-preserving abstraction on $\Prog$. The predicate transformer $\Wp_\Phi.\Prog$ is deterministic on cubed expectations.

\Proof
The result follows by showing that $\Wp.\Prog = \Wp_\Phi.\Prog$ on cubed expressions.\ Assume first that $c, c' \in \cubes.\Phi$, and that $\lambda, \lambda'$ are reals. We reason as follows:

\begin{Reason}
\Step{}
{\Wp_\Phi.\Prog.(\lambda[c] + \lambda'[c'])}
\StepR{$\leq$}{\Lem{l1541} (1)}
{\Wp.\Prog.(\lambda[c] + \lambda'[c'])}
\StepR{$=$}{$\Prog$ is deterministic}
{\lambda\Wp.\Prog.[c] + \lambda'\Wp.\Prog.[c']}
\StepR{$=$}{$\Prog$ is information-preserving}
{\lambda\Wp_\Phi.\Prog.[c] + \lambda'\Wp_\Phi.\Prog.[c']}
\StepR{$\leq$}{\Lem{l1541} (3)}
{\Wp_\Phi.\Prog.(\lambda[c] + \lambda'[c'])~.}
\end{Reason}
Observe finally that the equality generalises for expressions consisting of finite sums of cubes, and the fact that there are only finitely many distinct cubes whenever $\Phi$ is finite.
\end{Lems}

In particular we can now see that information-preserving abstractions compute exact results for all cubed expressions:

\begin{Corl}\label{c1233}
Let $\Prog$ be a deterministic (probabilistic) program, and let $\Phi$ be a set of predicates inducing an information-preserving abstraction on $\Prog$. Then $\Wp_\Phi.\Prog.e = \Wp.\Prog.e$ whenever $e$ is cubed.
\Proof
Follows since if $e$ is cubed then it is a finite sum of scaled cubes, and by \Lem{l0723} $\Wp_\Phi.\Prog$ distributes addition.
\end{Corl}

\subsection{Computing abstractions component-wise}

The above notions assume that the abstraction is calculated wholesale on the program $\Prog$; in practice it may be more efficient to calculate the abstraction by computing it relative to, and on smaller components of the program, however as \Lem{l1508} (5,6) indicate, additional inaccuracies can creep in wherever the abstraction is computed from program components.

Fortunately this does not occur in the case of information-preserving abstractions: \Lem{l0723} is key to verifying that information-preserving abstractions are determined from their components alone, provided that they themselves are also information-preserving. In practical terms this means that in a transition-system, provided each transition preserves the information, then so will the abstraction. In our predicate transformer framework, we need to show that $\Wp_\Phi$ distributes sequential composition and probabilistic choice.

\begin{Lems}\label{l1626}
Let $\Prog, \Prog'$ be deterministic (probabilistic) programs, and let $\Phi$ be a set of predicates inducing an information-preserving abstraction on each. The following inequalities apply:

\[
\begin{array}{llll}
(5')~~~ & \Wp_\Phi.\Prog . (\Wp_\Phi. \Prog')   &\Wide{=}& \Wp_\Phi.(\Prog ; \Prog')  \\

(6')~~~ &  \Wp_\Phi.\Prog \PC{p} \Wp_\Phi. \Prog'   &\Wide{=}& \Wp_\Phi.(\Prog \PC{p} \Prog') ~. \\
\end{array}
\]

\end{Lems}

\Proof
Follows easily from \Lem{l0723} since $\Wp_\Phi.\Prog$ and $\Wp_\Phi.\Prog'$ are both cubed expressions.


\subsection{Computing average performance}

Significantly, we can now compute expected performance profiles exactly from the abstraction.


\begin{Lems}\label{l0939}
Let $\Prog$ be a deterministic program, and information-preserving with respect to $\Phi$, and suppose that $G, (n\leq k) \in \Phi$, where $k$ is an integer.
The following equalities hold:
\[
\Wp.(\Do G \rightarrow \Prog; n\Gets n{+}1 \Od).[n\leq k] \Wide{=}  \Wp.(\Do \hat{G} \rightarrow \Prog_\Phi; n\Gets n{+}1 \Od).[n\leq k]~,
\]
where $\hat{G}$ represents the abstraction of $G$ in $S/\sim_\Phi$.

\Proof\\
Let $N \Defs \Wp.(\Do G \rightarrow \Prog; n\Gets n{+}1 \Od).[n\leq k]$, and $\mathord{N_\Phi \Defs  \Wp.(\Do \hat{G} \rightarrow \Prog_\Phi; n\Gets n{+}1 \Od).[n \leq k]}$.\ By \Lem{l1541} (1), and monotonicity of the programming language \figg{f0604} we see that $N_\Phi \leq N$. To show that $N\leq N_\Phi$ we note that $N$ and  $N_\Phi$ are both least fixed points of monotone expectation-to-expectation functions. We use the least fixed point property of functions over partially-ordered sets, \IE\ that if $f.x \leq x$ then $\mu.f \leq x$ \cite{Tarski}. Applied to $N$ and $N_\Phi$ we establish that $N_\Phi$ satisfies the least fixed point equation for $N$ as follows:

\begin{Reason}
\Step{}
{[G]{\times}[n\leq k] + [\neg G]{\times} \Wp_\Phi.\Prog.N}
\StepR{$=$}{$\cubed_\Phi.N_\Phi = N_\Phi$ (see below);  \Lem{l0723}}
{[G]{\times}[n\leq k] + [\neg G]{\times} \Wp.\Prog.N_\Phi}
\StepR{$=$}{$N_\Phi$ is a fixed point}
{N_\Phi~.}
\end{Reason}
The result now follows since $N$ is the least fixed point of the function $(\lambda x \cdot [G]{\times}n + [\neg G]{\times} \Wp.\Prog.x)$.

For the ``see below" part, note that $N_\Phi$ is itself a fixed point, satisfying: $N_\Phi = [G]{\times}[n\leq k] + [\neg G]{\times} \Wp_\Phi.\Prog.N_\Phi$. It now follows that $N_\Phi$ is cubed since  $\Wp_\Phi.\Prog.e$ is, for any expression $e$.
\end{Lems}

More generally exact bounds can be computed even when the program exhibits finitely-branching nondeterminism.

\begin{Corl}\label{c1021}
Let $\Prog_1\dots \Prog_m$ be deterministic and information-preserving with respect to $\Phi$. Let $G \in \Phi$, and $n$ a fresh variable. The following equality holds:

\[
\begin{array}{ll}
&\Wp. \Prog.(\Do G \rightarrow (\Prog_1 \Ch \dots \Ch \Prog_m \Od).[n\leq k]\\
=& \Wp. \Prog. (\Do G \rightarrow ({\Prog_1}_\Phi \Ch \dots \Ch{\Prog_m}_\Phi \Od).[n\leq k]~.\\
\end{array}
\]
\Proof
The proof is similar to \Lem{l0939} since nondeterminism distributes by \Lem{l1541} (4).
\end{Corl}

The significance of \Cor{c1021} is that whenever the abstraction is known to be information-preserving component-wise over a program (or transition system), then exact performance can be carried out on the abstraction. An important class of such programs are the so-called ``data independent" systems, to which we turn in the next section.

\subsection{Data independence}

A program is said to be \emph{data independent} (with respect to a data type $X$) \cite{Wolper86} if it cannot perform operations involving specific values of  the type: specifically it can only input, output, store and make comparisons using any relational operator $\Theta \in$ $\{=, <, \le, >, \ge, \dots\}$.\ Wolper points out that many distributed protocols fall into this category --- he shows that such systems can be model checked accurately.  In fact, if we extend this informal definition to probabilistic programs such that all probabilistic choices are constants, then our results above imply that there is an abstraction which can be used to compute performance properties exactly, namely the abstraction induced by predicates $\Psi \Defs \{x \ \Theta \ y \ | \ \forall x,y \ $program \ variables of same type$ \}$.\ We use this intuition to define a simple characterisation of data independent programs: they are the programs which are information-preserving with respect to  $\Psi$ (with informal definition above).

 \begin{Defns}\label{d1100}
 Let $\Prog$ be a deterministic $\pGCL$ program with variables $x_1 \dots x_m$. We say that $\Prog$ is \emph{data independent} with respect to $x_1 \dots x_m$ provided that $\Prog$ is information-preserving with respect to the abstraction induced by $\Psi$, where $\Psi$ is the set of predicates containing all expressions of the form $x_{i} \ \Theta \ y_{j}$ for all $1\leq i, j, \leq m$.
 \end{Defns}

\noindent Note that this characterisation of data independence can be generalised to programs $\mathord{\Prog_1 \Ch \dots \Ch \Prog_n}$ which exhibit nondeterministic behaviour by ensuring that the deterministic components $\Prog_i$ comply with \Def{d1100}. Note that this definition shares some similarities with Wolper's denotational characterisation \cite{Wolper86}, in that \Def{d1100} captures the idea that properties expressible at the granularity of $\Phi$ are shared by both $\Prog_\Phi$ and $\Prog$.  It does not deal with general types however, as does Lazic \cite{Lazic99}.

 With \Def{d1100} we can now conclude that data independent probabilistic programs have abstractions which preserve performance bounds.%

 \begin{Lems}\label{l1111}
 Let $\Prog$ be a data independent program. Then the expected number of iterations $\Do G \rightarrow \Prog  \Od$ may be computed exactly using the abstract program $\Prog_\Psi$ whenever $G \in \Psi$, where $\Psi$ is defined in \Def{d1100}.
 \end{Lems}

 The practical implication of \Lem{l1111} (which follows directly from \Lem{l0939} and \Cor{c1021}) is that performance (and correctness) of data independent programs can be \emph{analysed exactly} using  model checking.\ In the next section we give an example to illustrate this idea.

\section{Case study: Rabin's distributed consensus}\label{DC}

We illustrate the effectiveness of our technique on
the Rabin's choice-coordination problem \cite{Rabin82}.\ The state space generated on execution of the algorithm is potentially infinite hence limiting the scope of model checking on verifying liveness properties (such as termination conditions) relating to its overall performance.\ As we will see, even though the algorithm is not quite data independent, there does exist an information-preserving abstraction demonstrating that exact numerical analysis is still possible on its performance.

\subsection{Informal description}
A group of tourists are to decide between two meeting places (which are not of
interest to us).\ A major constraint is that they may not communicate as a group; nor
is there a central ``authority'' ($e.g.$ a tour guide) whose decision overrides theirs.

Each tourist carries a notepad on which he will write various numbers; outside each of
the two potential meeting places is a noticeboard on which various messages will
be written.\ Initially the number zero appears on all the notepads and on the two
noticeboards.

Each tourist decides independently (demonically) which meeting place to visit first, after
which he strictly alternates his visits between them.\ At each place he looks at the
noticeboard, and if it displays ``here'', he goes inside.\ If it does not display ``here''
it will display a number instead, in which case the tourist compares that number $K$ with
the number $k$ on his notepad and does one of the following:

\begin{itemize}
\item []if $k < K$ --- The tourist writes $K$ on his notepad (erasing $k$), and goes to the
        other place.
\item [] if $k > K$ --- The tourist writes ``here'' on the noticeboard (erasing $K$), and
        goes inside.
\item [] if $k = K$ --- The tourist chooses $K^{'}$, the next even number larger than $K$,
        and then flips a coin: if it comes up heads, he increases $K^{'}$ by a further one.\
        He then writes $K^{'}$ on the noticeboard and on his notepad (erasing $k$ and $K$),
        and goes to the other place.\footnote{For example if $K$ is 2 or 3, then $K^{'}$ becomes
        4 and then possibly 5.}
\end{itemize}
A key characterisation of the Rabin's algorithm, which has also been proved elsewhere \cite{ARP} is that, on termination all the tourists' will converge at the same meeting place, and that happens with probability 1.\ However it is not always the case that an ``observer'' can witness every state of the program that will
lead to termination.\ For example, it is possible that {\it the tourists will forever (according to
an observer) keep updating their notepads and the noticeboards without deciding on a meeting
place.}\ This enforces an {unbounded state} behaviour on the algorithm.\ Nonetheless, given the unbounded state nature of the algorithm,
our theoretical results still permit us to study a suitable performance attribute of the system: the expected number of rounds (or steps) of the protocol until termination (analogous to convergence).



\subsection{A pGCL snapshot of the Rabin's algorithm}
Fig. {\ref{fig:DCpGCL}} (on the next page) gives an overview of the Rabin's choice-coordination
problem in the pGCL.\ We call the two meeting places ``left'' and ``right'' as we discuss it and refer to the
notations\footnote{$\bag{...}{}$ --- bag (multiset) brackets; $\square$ ---empty bag; $\bag{n}{N}$---bag containing $N$ copies all of value $n$; {\bf take} $n$ {\bf from} $b$ --- a program command which chooses an element demonically from non-empty bag $b$, assigns it to $n$ and removes it from $b$; {\bf add} $n$ {\bf to} $b$ ---add element $n$ to bag $b$; $\overline{n}$ ---the ``conjugate'' of $n$, it is $n + 1$ if $n$ is even and $n - 1$ if $n$ is odd; $\#b$ --- the number of elements in a bag b.} accordingly.\
Bag {\it lout (rout)} is the bag of numbers held by tourists waiting to look at the left (right) noticeboard;
bag {\it lin (rin)} is the bag of numbers held by tourists who have already decided on the left (right)
alternative; number {\it L (R)} is the number on the left (right) noticeboard.\ Initially there are {\it A (B)} tourists
on the left (right), all holding the number zero; no tourist has yet made a decision, and both noticeboards
show zero.

Execution is as follows: if some tourists are still undecided (so that {\it lout (rout)} is not yet empty), select one:
the number he holds is {\it l (r)}.\ If some tourist has already decided on this alternative (so that {\it lin (rin)} is
not empty), this tourist does the same; otherwise any of the three possibilities discussed above
is executed.

\begin{figure}
\small
\[
\begin{array}{l}
lout, rout \ \Defs \
\hspace*{-1ex}
\ \bag{0}{A},\ \bag{0}{B} ; \\
lin, rin \ \Defs \
\hspace*{-1ex}
\ \square,\ \square \ ; \\
L, R \ \Defs \
\hspace*{-1ex}
\ 0,\ 0 \ ; \\
\\
\hspace*{-1ex}\begin{array}[t]{l}
{\bf do}\ \ lout \neq \square \rightarrow \\

\quad\quad ${\bf take}$ \ l \ ${\bf from}$ \ lout \ ; \\
\quad\quad {\bf if} \ lin \neq \square \ {\bf then} \ {\bf add} \ l \ {\bf to} \ lin \ {\bf else}\\
\quad\quad\quad l > L \rightarrow \ {\bf add} \ l \ {\bf to} \  lin\\
\quad\quad []\ l = L \rightarrow \ (L:= L + 2 \pp{\frac{1}{2}} \overline{(L + 2)} ; \ {\bf add} \ L \ {\bf to} \ rout\\
\quad\quad []\ l < L \rightarrow \ {\bf add} \ L \ {\bf to} \ rout\\
\quad\quad {\bf fi}\\
\\
$[]$ \ \ rout \neq \square \rightarrow \\
\quad\quad ${\bf take}$ \ r \ ${\bf from}$ \ rout \ ; \\
\quad\quad {\bf if} \ rin \neq \square \ {\bf then} \ {\bf add} \ r \ {\bf to} \ rin \ {\bf else}\\
\quad\quad\quad r > R \rightarrow \ {\bf add} \ r \ {\bf to} \  rin\\
\quad\quad []\ r = R \rightarrow \ (R:= R + 2 \pp{\frac{1}{2}} \overline{(R + 2)} ; \ {\bf add} \ R \ {\bf to} \ lout\\
\quad\quad []\ r < R \rightarrow \ {\bf add} \ R \ {\bf to} \ lout\\
\quad\quad {\bf fi}\\
{\bf od}
\end{array}
\end{array}
\]
\caption{\rm The Rabin's choice coordination algorithm in the pGCL (adapted from \cite{ARP}).}\label{fig:DCpGCL}
\end{figure}

\subsection{Computing average performance of the Rabin's algorithm}
In this section we discuss properties of the Rabin's algorithm sufficient for an analysis of its average performance.\ Since the unbounded state nature of the algorithm limits the scope of model checking on the original system, we must therefore compute a suitable abstraction prior to performance analysis.\ Nevertheless, with our proposed technique, it is possible to defeat the overhead incurred by model checking the unbounded state system just by model checking its information-preserving abstraction.

One performance property of interest is captured by computing the minimum probability $Pmin$, that within a finite number of  steps $T$, the tourists eventually converge at the same meeting place on termination.\ In the logic PCTL \cite{PCTL}, directly supported by the PRISM tool, we express this property as
\begin{eqnarray}\label{criterion}
\quad Pmin=? [true  \ U^{\le T} \ (\#lin = N) \ | \ (\#rin = N)],
\end{eqnarray}
where $N$ represents the total number of tourists who will initially decide on where to meet {\it i.e.} $\mathord{N = A + B}$.\

Similarly, with the reward structures \cite{KNP07} of the PCTL we compute the expected number of rounds of the protocol until termination.\ Again, this will be done on the protocol's abstraction using the specification:
 \begin{eqnarray}\label{reward}
\quad Rmin \ | \ Rmax =? [F \ (\#lin = N) \ | \ (\#rin = N)].
\end{eqnarray}
The parameters $Rmin$ and $Rmax$ respectively represent the expected minimum and maximum rewards (expected number of rounds)
until the tourists eventually converge at the same meeting place.\ We note that states where the tourists have not yet
 met the convergence condition are worth a reward value of one.

In the sections that follow, we explain how we identify essential behaviours of the algorithm that will permit the construction of an information-preserving abstraction upon which the analysis can be performed.

\subsection{An information-preserving abstraction}
As earlier stated, even though the Rabin's algorithm has unbounded state behaviour, it is not data independent since the probabilistic update increments the variables $L,R$ {\it etc}.\ However there is still an information-preserving abstraction, which we will now describe.

Observe that although the noticeboard values are incremented, they always maintain $|L-R| \leq 2$.\ In terms of the algorithm, the only information that needs to be preserved is the value $L-R$ and whether $L,R$ are odd or even.\ Finally, the relative values of the tourists' numbers to $L$ and $R$ also need to be recorded, as well as their location.\ This generates an information-preserving abstraction.\ In practice, we characterise the relationship between the noticeboard values using a fresh variable we call {\it slot}, which can only take values in $\{$0, 1, 2$\}$ --- since the noticeboard values and hence the notepad values can only lie in one of these slots for any given state of the system.\ We define the slot variable as follows and interpret transitions in the abstract state with respect to the slot values:

\[ slot \Defs \left\{ \begin{array}{l}
 \hspace*{-1ex}\begin{array}[t]{l}
\ 0 \quad \quad $if$ \quad L = R \\
\ 1 \quad \quad $if$ \quad  L = R - 2\  \vee \ R = L - 2\\
\ 2 \quad \quad $if$ \quad  L = \overline{R}\ \ .
\
\end{array}
\end{array}\right.\]

\begin{figure}
\begin{center}
\includegraphics[scale = 0.45]{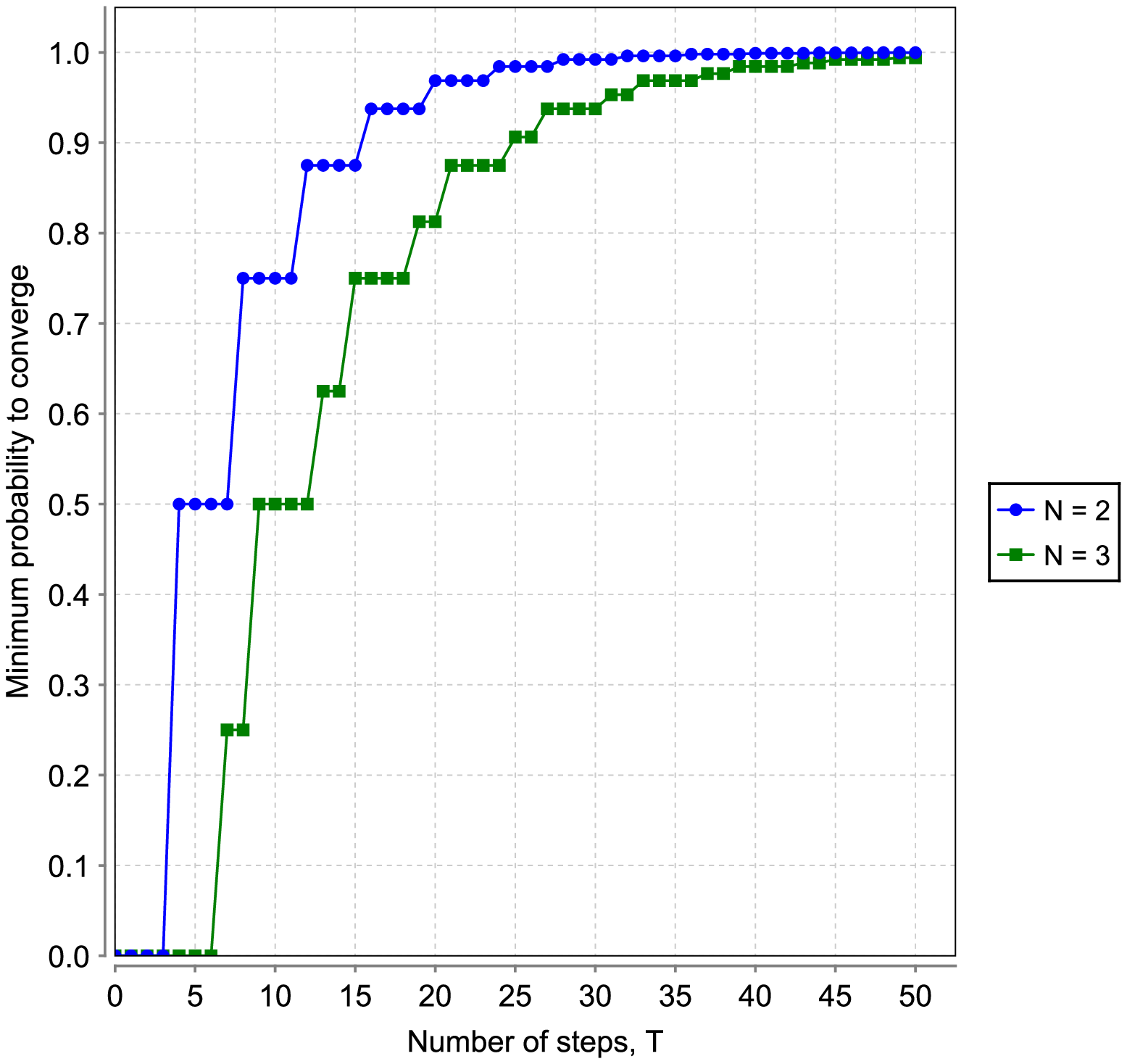}
\end{center}\vspace{-0.5cm}

\caption{Performance result of Rabin's algorithm for N = 2, 3.}\label{fig:abstraction}
\end{figure}
In the section that follows, we explain the performance results derived from the information-preserving abstraction.\ The results nevertheless give a precise summary of the performance of the original system.

\begin{figure}
\begin{center}
\noindent\begin{tabular}{|c|c|c|}
\hline
{\quad \bf Number of Tourists\quad } & {\quad \bf Rmin \quad } & {\quad \bf Rmax \quad}\\ \hline
{\bf N = 2} &  2 & 7 \\
\hline
{\bf N = 3} &  2 & 11 \\
\hline
\end{tabular}
\caption{\rm Expected number of steps} \label{fig:analysis}\vspace{-.5cm}
\end{center}
\end{figure}

\subsection{Experimental results}
We model the abstract behaviour discussed above for the base cases of even and odd number of tourists ($N = 2,3$) in the PRISM language, and similarly analyse the performance results as captured by the properties in (\ref{criterion}) and (\ref{reward}), using the experimentation facility of the tool.\ A similar model construction and analysis for larger values of $N$ is also possible by repeating the same technique although very laborious.

\figg{fig:abstraction} captures the performance characteristics of the information-preserving abstraction of the Rabin's algorithm.\ It clearly establishes the termination property of the unbounded state system using just its abstraction: note that both graphs converge to probability 1.\ In the original unbounded state system, achieving this is practically impossible.\ See the original model in the compendium of case studies at \cite{PRISM}.

We also observe (\figg{fig:analysis}) that the expected minimum and maximum number of rounds until termination can be model checked, and hence nevertheless gives an exact bound on the number of steps required for the unbounded state system to terminate.\ Again, in the unbounded state system, the result of computing $Rmax$ for example is {\it infinity}, which in the PRISM tool is interpreted to mean that it is not possible for a terminating (or convergence) condition to be reached.

\section{Discussion}

While some probabilistic program logics allow programs to be compared even at abstract levels, for example using the techniques
in \cite{LS91, JS90}, the underlying logic of the pGCL supports the notion of program refinement and hence compositionality.\ This
 makes it easy to relate refinement over concrete states to their abstract counterparts and furthermore with the other 
 probabilistic program logics, given any context.

Other approaches seek to use variations of counterexample guided predicate abstraction \cite{VMCAI,CEGAR} to automate finding sets of predicates which generate finer abstractions.\ One way to see the relationship with our approach would be to note that when an abstraction is observed to be
information-preserving (according to \Lem{l1307} for example) then further refinement is unnecessary.\ Kwiatkowska et al. \cite{QEST} propose an approach to estimate the accuracy of the analysis implied by any abstraction, confirming that for information-preserving abstractions the analysis is exact.

On the application level, one way to see the usefulness of our technique is in the recent research direction of linking proof-based verification with model checking for probabilistic systems \cite{Ndukwu09, Ndukwu10}.\ Since proof-based verification can cope with proofs over infinite state systems, a key challenge with this technique is then the identification and constructing of information-preserving abstractions upon which a model checking algorithmic verification can be performed.\ This is still an open problem.

\section{Conclusion and future work}

In this paper we have developed the theory of predicate abstraction for probabilistic programs within the framework of expectation transformers.\ We have similarly established a criterion to help discover when abstractions do not lose information especially for probabilistic programs; and we have demonstrated the applicability of the results to data independent programs (or at least their approximations).

Whilst our theoretical approach identifies when a set of predicates is information-preserving, it does not provide assistance for finding one.\ Even though we have computed the abstraction by hand, we quickly remark that applying the manual construction technique for $N > 3$ would seem a laborious task.\ Note that our technique results in a huge success for verifying the termination condition of the algorithm when compared with the concrete system as modeled in the compendium of case studies at the URL \cite{PRISM}.

However, a future direction for this work would be to develop an automated strategy which would construct abstractions ``on the fly'', given that our theoretical framework is rich enough to provide intuitions to identifying sets of suitable predicates to aid the construction of information-preserving abstractions.\\\\
{\bf \noindent \large{Acknowledgement:}} The authors are grateful to the anonymous reviewers for their helpful comments.

\bibliography{DataIndependence}
\bibliographystyle{plain}


\end{document}